\documentclass[twocolumn, twocolappendix]{aastex631}

\usepackage{amssymb}
\usepackage{amsmath}
\usepackage{soul}
\newcommand{\be}{\begin{equation}}
\newcommand{\ee}{\end{equation}}
\newcommand{\R}[1]{\textcolor{red}{#1}}
\newcommand{\B}[1]{\textcolor{blue}{#1}}

\newcommand{\vect}[1]{\boldsymbol{#1}}

\shorttitle{Uncertainty and bias of dark sirens}
\shortauthors{Yu, Seymour, Wang, \& Chen}
\begin{document}

\title{Uncertainty and bias of cosmology and astrophysical population model from statistical dark sirens %via the Fisher information approach
}

\correspondingauthor{Hang Yu}
\email{hyu45jhu@gmail.com}

\author[0000-0002-6011-6190]{Hang Yu}
\affiliation{TAPIR, Walter Burke Institute for Theoretical Physics, Mailcode 350-17\\ California Institute of Technology, Pasadena, CA 91125, USA}

\author[0000-0002-7865-1052]{Brian Seymour}
\affiliation{TAPIR, Walter Burke Institute for Theoretical Physics, Mailcode 350-17\\ California Institute of Technology, Pasadena, CA 91125, USA}

\author[0000-0002-5581-2001]{Yijun Wang}
\affiliation{TAPIR, Walter Burke Institute for Theoretical Physics, Mailcode 350-17\\ California Institute of Technology, Pasadena, CA 91125, USA}

\author{Yanbei Chen}
\affiliation{TAPIR, Walter Burke Institute for Theoretical Physics, Mailcode 350-17\\ California Institute of Technology, Pasadena, CA 91125, USA}




\begin{abstract}
% 1. statistical dark siren is a crucial way for us to constrain cosmology. \\
% 2. it can be viewed as a comparison between the observed event distribution and the expected one based on astrophysical model. \\
% 3. uncertainties on the cosmological and astrophysical parameters can be obtained from the Fisher information on the histograms. \\
% 4. enables a simple and fast way to estimate the constraints to be placed by future detectors. \\
% 5. gain insights in the calculations. \\
% 5. Using this method, we consider the bias due to substructures in the merger rate and population mass distribution. We find a $1\%$ deviation in the astrophysical model can lead to $1\%$ error in the Hubble constant.

Gravitational-wave (GW) radiation from a coalescing compact binary is a standard siren as the luminosity distance of each event can be directly measured from the amplitude of the signal. 
% While without the assistance of an electromagnetic counterpart, the redshift of a single binary black hole (BBH, which corresponds to a dark siren) event is perfectly degenerate with its mass, from a population of events the redshift distribution can nonetheless be constrained, thereby allowing us to statistically measure cosmological parameters including the Hubble constant.
% While we cannot directly measure cosmological parameters from a single dark siren event (i.e., an event without an electromagnetic counterpart) due to the degeneracy between mass and redshift, we can nonetheless statistically infer cosmology from a population of events.  
One possibility to constrain cosmology using the GW siren is to perform statistical inference on a population of binary black hole (BBH) events.  
In essence, this statistical method can be viewed as follows. 
We can modify the shape of the distribution of observed BBH events by changing cosmological parameters until it eventually matches the distribution constructed from an astrophysical population model, thereby allowing us to determine the cosmological parameters.
In this work, we derive the Cram\'er-Rao bound for both cosmological parameters and those governing the astrophysical population model from this statistical dark siren method by examining the Fisher information contained in the event distribution. 
Our study provides analytical insights and enables fast yet accurate estimations of the statistical accuracy of dark siren cosmology. 
Furthermore, we consider the bias in cosmology due to unmodeled substructures in the merger rate and the mass distribution. We find a $1\%$ deviation in the astrophysical model can lead to a more than $1\%$ error in the Hubble constant. This could limit the accuracy of dark siren cosmology when there are more than $10^4$ BBH events detected. 
\end{abstract}


\section{Introduction}
\label{sec:intro}

%GW are standard sirens that may resolve the tension between CMB and type-Ia. 
The key to study modern cosmology is to measure a relation between distance and redshift. In electromagnetic (EM) observations, the redshift to the source can be directly measured (e.g., by comparing the measured spectra to the ones obtained in terrestrial laboratories), and the challenge is to constrain the distance. To do so, it relies on utilizing some forms of standard references. One possibility is to use ``standard candles'' with known intrinsic luminosity, and the best-known example is a type-Ia supernovae~\citep{Riess:96, Riess:21}. Another possibility is to use a ``standard ruler'' with a known size, and the imprint of sound waves in the Cosmic Microwave Background is such an example~\citep{Spergel:03, Planck:13, Planck:18}. However, a tension on the value of the Hubble Constant, conventionally denoted by $H_0$, emerges between the latest results of the two sets of measurements~\citep{Verde:19}. It thus calls for a third method to either reconcile or confirm the tension. 

This brings observations using gravitational waves (GWs) to people's attention, a new possibility opened up by Advanced LIGO (aLIGO; \citealt{LSC:15}), Advanced Virgo~\citep{Virgo:15}, and KAGRA~\citep{KAGRA:18, KAGRA:21}. GW events are ``standard sirens'' in cosmology~\citep{Schutz:86, Holz:05} as the amplitude of an event encodes directly the luminosity distance to the source. If the redshift information can be further constrained, we can then determine the values of cosmological parameters.

One way to obtain the redshift information is through multi-messenger observation of an event. If we can simultaneously observe a GW event and its EM counterpart, corresponding to a ``bright siren'', we can then identify the host galaxy of the event, from which we can further extract the redshift~\citep{Holz:05, Chen:18}. A GW event involving neutron stars (either a binary neutron star, or BNS, or a neutron star-black hole event) is an ideal candidate here. Indeed, the first BNS event, GW170817, is a highly successful example~\citep{LVC:17bnsMultiMess, LVC:17bnsH0}. From this event alone, we were able to constrain the Hubble constant to $H_0=70_{-8}^{+12}\,{\rm km\,s^{-1}\,Mpc^{-1}}$ within the 68\% credible interval. With future detectors like LIGO-Voyager~\citep{Adhikari:20} or third-generation (3G) GW detectors including the Einstein Telescope~\citep{Sathyaprakash:11} and the Cosmic Explorer~\citep{Evans:17, Reitze:19, Evans:21}, it is potentially possible to constrain $H_0$ with percent level accuracy and the normalized matter density $\Omega_m$ to an accuracy of $\mathcal{O}(10\%)$~\citep{Chen:21}. 
However, such bright sirens are rare and GW170817 is the only joint observation to date. Even with 3G detectors, \citet{Califano:22} estimates that only $0.1\%$ of detectable BNSs will have observable EM counterparts. 
Besides a direct EM counterpart, it is also possible to constrain cosmology from matter effects in coalescing BNSs~\citep{Messenger:12}. 
% \bs{BNS hubble constant measurements also possible with tidal effects for non EM cases \cite{Messenger:2011gi}.}
% \ali{suggestion: For Einstein Telescope, it is estimated that approximately $0.1\%$ of all detectable BNSs have observable EM counterparts \cite{Califano:22}}

Alternatively, we may further utilize information in binary black hole (BBH) events, which consists of the majority of event catalogs \citep{LVC:16pop, LVC:19gwtc1, LVC:21gwtc2, LVC:21gwtc2.1, LVK:21gwtc3, Nitz:19, Nitz:20, Gray:2019ksv, Venumadhav:20, Olsen:22}. An EM counterpart is typically not expected for a BBH event, and therefore a BBH corresponds to a dark siren (though a counterpart might be possible if the BBH resides in a gaseous environment; see, e.g., \citealt{McKernan:19}). While for a single event, it is challenging to obtain the redshift due to the perfect degeneracy between redshift and mass (unless the source can be accurately localized to only a few potential host galaxies, a point we will get back to at Sec.~\ref{sec:conclusion_discussion}), we can nonetheless infer the redshift distribution of a collection of BBH events statistically. 
% \ali{paraphraze: we can infer the redshift distribution of a collection of events..?}

Initially, the statistical inference was done by comparing a BBH event catalog with galaxy catalogs (e.g., \citealt{Schutz:86, Chen:18, Fishbach:19, Finke:21}). Later, people realized that features in the mass distribution of BBH events could also be used to constrain the cosmological parameters (e.g., ~\citealt{Chernoff:93, Taylor:12, Farr:19, Mastrogiovanni:21, LVK:21cos, MariaEzquiaga:22}). 
In both cases,  one computes the likelihood of each event to happen given a set of cosmological parameters as well as an assumed astrophysical population model. The likelihood for all the events are then multiplied together to get the likelihood of the observed population given the assumed cosmological and astrophysical parameters. This is further converted to a posterior distribution of parameters with an assumed prior distribution~\citep{Mandel:19, Thrane:19}. 

In essence, the statistical approach corresponds to a comparison between two histograms, or distributions. One distribution is obtained from the observed BBH events with respect to either the luminosity distance or detector-frame masses (or both as a high-dimensional distribution). The other distribution is constructed from our astrophysical model with respect to either redshift or source-frame masses (or both). By varying the values of cosmological parameters, as well as those governing the astrophysical population, we can eventually match up the two distributions, thereby constraining cosmology and population model simultaneously. %\ali{paraphrase: constraints on the cosmology and the population model follow from demanding consistency between these two distributions.}\hang{(HY: I would slightly prefer the original way to emphasize the variation in the distribution caused by varying cosmological parameters.)}
%One from GW events and one from (model-dependent) astrophysical expectation. 

With this view, we propose an especially convenient way to assess the statistical power of dark siren cosmology. In particular, we can analytically construct the Fisher information encoded in the distributions. From that, we can both estimate the uncertainties on the parameters governing the distributions and understand correlations among the parameters. As we will show later, even with a few simplifying assumptions, this approach predicts a similar level of uncertainty on the Hubble constant when applied to the GWTC-3 catalog~\citep{LVK:21gwtc3}, as well as many other key features obtained in \citet{LVK:21cos}. It also reproduces the results of previous studies (e.g., \citealt{Fishbach:18, Farr:19}) when forecasting the future constraints on both the population model and cosmology with hundreds to thousands of BBH events. Therefore, our approach serves as a simple and analytical way to study the statistical dark siren method, which can be especially useful when making quick but decently accurate predictions for the future when a large number of events are expected. It thus complements the more accurate yet also more complicated hierarchical inference approach~\citep{Mandel:19}. 

Furthermore, our approach can be used to study the bias on cosmological and/or astrophysical parameters due to errors in the assumed population models. 
We will first provide a general framework to study the bias due to any form of errors, and then as a case study, we fill examine in detail how unmodeled substructures in the mass and/or redshift model would affect the inference of the Hubble constant. 
This is motivated by the latest population model by \citet{LVK:21pop} where signs of substructures are suggested. 
%Signs of sub-structures are already suggested by the GWTC3 population model. 

The rest of the paper is organized as follows. In Sec.~\ref{sec:framework}, we provide the mathematical framework to construct the Fisher information matrix of a distribution, which estimates the covariance matrix when jointly fitting cosmological parameters and population properties. We will also consider the bias induced on the cosmological parameters due to structures not captured by a parameterized population model with a specific functional form. We then describe the astrophysical model adopted in our study in Sec.~\ref{sec:model}. The application to the GWTC-3 catalog is presented in Sec.~\ref{sec:GWTC3}. To further validate our method, we also present the reproduction of previous studies' results using our method in App.~\ref{sec:validation}. In Sec.~\ref{sec:bias}, we consider the bias on cosmological inference induced by unmodeled substructures in both the mass distribution and merger rate function, and we set requirements on the accuracy of the population in order for the bias to be below the statistical error. Lastly, we conclude and discuss in Sec.~\ref{sec:conclusion_discussion}.

\section{Basic framework}
\label{sec:framework}

\begin{figure}
  \centering
  \includegraphics[width=\columnwidth]{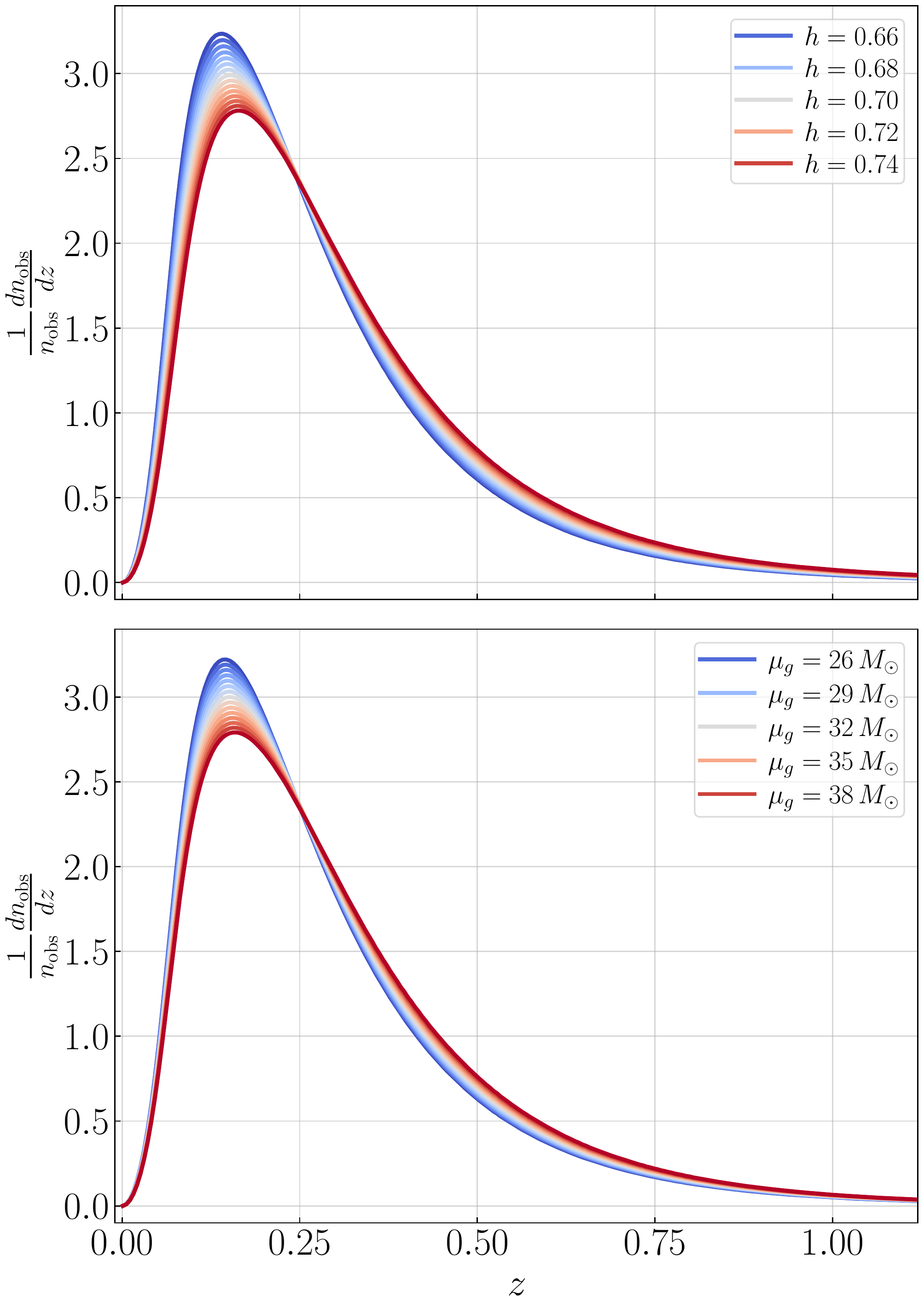}
  \caption{Top: expected number of detection as a function of the redshift $z$ at different values of $h\equiv H_0/(100\,{\rm km/s/Mpc})$. From GW events, we can construct such an distribution as a function of $D_L$ first and then convert it to a function of $z$ based on assumed cosmological parameters. Meanwhile, our astrophysical knowledge allows us to construct an expected distribution as a function of $z$ from, e.g., galaxy catalogs. By comparing the two histograms, we can then constrain the value of cosmological parameters. 
  Bottom: the distribution is also affected by astrophysical models (e.g., the location of a peak in the BBH's mass distribution $\mu_g$; see Sec.~\ref{sec:model}) which could mimic the effect of changing cosmological model. This indicates the significance of jointly analyzing astrophysical and cosmological parameters. 
  }
\label{fig:sample_dist_vs_z}
\end{figure}

\begin{figure}
  \centering
  \includegraphics[width=\columnwidth]{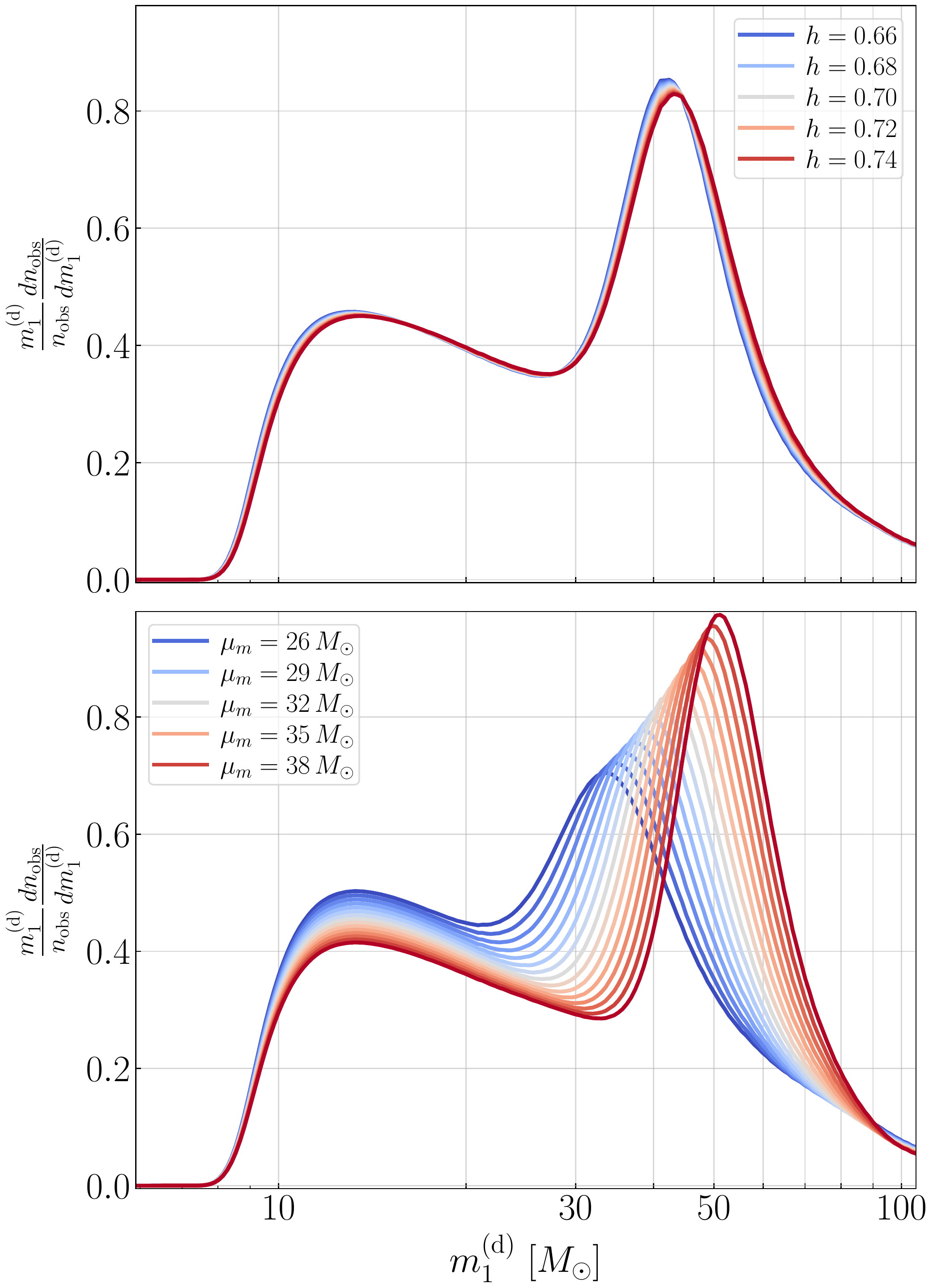}
  \caption{Similar to Fig.~\ref{fig:sample_dist_vs_z} but now we plot the distribution as a function of the detector-frame mass of the primary BH, $m_1^{\rm d}$. Combining it with Fig.~\ref{fig:sample_dist_vs_z} can thus be used to break the degeneracy between astrophysical and cosmological parameters.
  }
\label{fig:sample_dist_vs_m1det}
\end{figure}

We demonstrate in this work that in essence, the statistical dark siren approach corresponds to a comparison between a measured distribution of GW events and the one we construct based on our knowledge (or assumption) of the cosmology and the astrophysical source population. 

Examples are illustrated in Figs.~\ref{fig:sample_dist_vs_z} and \ref{fig:sample_dist_vs_m1det}. Here the y-axis is the normalized detection probability density of GW events (the parameters are consistent with those inferred from GWTC-3; Sec.~\ref{sec:GWTC3}).
The x-axis can be the redshift $z$ or the mass of the primary (either the detector-frame one $m_1^{\rm (d)}$ or the source-frame one $m_1$). While for illustration purpose we focus on marginalized one-dimensional distributions, the analysis in this section can be straightforwardly extended to high-dimensional distributions as well. 

Without loss of generality, we can construct a histogram of observed BBHs events with respect to a general coordinate $x$ (which can be the redshift $z$, the mass of the primary black hole $m_1$, or other quantities). The expected number of observations in the $i$'th bin at $[x_i, x_i{+}\Delta x)$ can be written as $r_i(\vect{\theta}^C, \vect{\theta}^A)\Delta x$, where $r$ is the event density. We use $\vect{\theta}^C=(H_0, \Omega_m, ...)$ to denote the cosmological parameters and $\vect{\theta}^A$ the other astrophysical parameters. The number of observations in the $i$'th bin, $n_i$, follows a Poisson distribution~\footnote{Here for simplicity, we ignore the inference uncertainty of each individual event's parameters (e.g., redshift and mass, etc.). As we will see in later sections, the results we obtain under this simplification is decently accurate. The uncertainty on individual event's parameters smears out fine details but keeps the broad, coarse-grained features in the population distribution. Current analysis focuses on the coarse-grained part (see, e.g., \citealt{LVK:21cos}), though for high-precision  cosmology, it would be critical to also capture substructures in the model (see later in Sec.~\ref{sec:bias}). 
%is likely averaged out when considering the distribution of the population. 
%\bs{What do you mean by "qualitatively accurate as shown in the later sections"? Are you just saying that our method replicates Bayesian methods?} 
A more general treatment incorporating the uncertainty (and potentially systematic bias) on individual events is deferred to a future study.}
\begin{equation}
    p\left[n_i|r_i(\vect{\theta}^C, \vect{\theta}^A)\right]= \frac{(r_i\Delta x)^{n_i}\exp(-r_i\Delta x)}{n_i!}.
\end{equation}

The Fisher information of $\vect{\theta}=(\vect{\theta}^C, \vect{\theta}^A)$ at a given bin $i$ is given by 
\begin{align}
    I_{i, ab} &= \sum_{n_i=0}^{\infty}p[n_i|r_i(\vect{\theta})] \left[\frac{\partial \log p(n_i|r_i)}{\partial r_i}\right]^2\left[\frac{\partial r_i(\vect{\theta})}{\partial \theta_a}\right]\left[\frac{\partial r_i(\vect{\theta})}{\partial \theta_b}\right], \nonumber\\
    &=
    \left[\frac{\partial r_i(\vect{\theta})}{\partial \theta_a}\right]
    \left[\frac{\partial r_i(\vect{\theta})}{\partial \theta_b}\right]
    \sum_{n_i=0}^{\infty}p[n_i|r_i(\vect{\theta})] \left[\frac{\partial \log p(n_i|r_i)}{\partial r_i}\right]^2, \nonumber \\
    &=\frac{\Delta x}{r_i}
    \left[\frac{\partial r_i(\vect{\theta})}{\partial \theta_a}\right]
    \left[\frac{\partial r_i(\vect{\theta})}{\partial \theta_b}\right],\nonumber \\
    &= r_i 
    \left[\frac{\partial \log r_i(\vect{\theta})}{\partial \theta_a}\right]
    \left[\frac{\partial \log r_i(\vect{\theta})}{\partial \theta_b}\right]
    \Delta x,
\end{align}
where we have used the subscripts $a,b$ to denote the $(a, b)$'th element in the Fisher information matrix and the derivatives are evaluated at the true values of $\vect{\theta}$ (or in practice, our best estimation of $\vect{\theta}$). 
Summing over all the bins and convert the discrete sum into an integral over $dx$, we thus arrive at the Fisher information matrix
\begin{equation}
    I_{ab}\left(\vect{\theta}\right) = \int r(x|\vect{\theta}) 
    \left[\frac{\partial \log r(x|\vect{\theta})}{\partial \theta_a}\right]
    \left[\frac{\partial \log r(x|\vect{\theta})}{\partial \theta_b}\right]
     dx.
     \label{eq:fisher}
\end{equation}
From the distribution, the covariance matrix of $\vect{\theta}$, $\vect{\rm Cov}(\vect{\theta})$, can be estimated by the Cram\'er-Rao bound as 
\begin{equation}
    \vect{\rm Cov}\left(\vect{\theta}\right) = \left[\vect{I}\left(\vect{\theta}\right)\right]^{-1}. 
\end{equation}

For future convenience, we also define $\vect{I}^C$ where the differentiation in Eq.~(\ref{eq:fisher}) is done only with respect to $\vect{\theta}^C$, or $\theta_{a,b}\in \left\{\vect{\theta}^C\right\}$. 
Effectively, $\vect{I}^C$ corresponds to the case where we have perfect knowledge on the astrophysical event rate, while $\vect{I}$ considers further the covariance between  astrophysical population models and cosmological parameters. 

Note that in the analysis above, we have assumed that the astrophysical model has the correct functional form and only has unknown parameter values. 
It might also be possible that the astrophysical model is formally inaccurate (e.g., due to substructures in the model and/or evolution in the population). In this case, the estimation of cosmological parameters can be systematically biased. 

To calculate the bias, we suppose the true rate (denoted by a superscript ``$t$'') in the $i$'th bin can be written as 
\begin{equation}
    r^t_i = r_i\left(\vect{\theta}^C, \vect{\theta}^A\right) + \Delta r_i
\end{equation}
% we change the notation slightly to write the probability at each bin as $p[n_i| \vect{\theta}^C, r_i(\vect{\theta}^A)]$.
We can expand the log-likelihood around the true $\vect{\theta}^C$ and $\Delta r_i=0$ (the expansion around $\vect{\theta}^A$ can be straightforwardly included; yet the covariance between $\vect{\theta}^A$ and $\vect{\theta}^C$ has been accounted for the Fisher matrix in Eq.~(\ref{eq:fisher}) and therefore we ignore it here),
\begin{align}
    \Delta \log p =& \frac{\partial^2 \log p}{\partial \theta^C_a \partial \theta^C_b}\Delta \theta^C_a\Delta \theta^C_b 
    +\frac{\partial^2 \log p}{\partial \theta_a^C\partial r_i^t} \Delta \theta^C_a \Delta r_i\nonumber \\
    +& \frac{\partial^2 \log p}{\partial \left(r_i^t\right)^2} \Delta r_i^2,
\end{align}
where the first derivative vanishes because at true values the probability is maximized. 

The bias in the cosmological parameter induced by $\Delta r_i$ is then given by setting
\begin{equation}
    0= \frac{\Delta \log p}{\Delta \theta_a^C},
\end{equation}
or
\begin{equation}
    \frac{\partial^2 \log p}{\partial \theta^C_a \partial \theta^C_b}\Delta  \theta^C_b 
    =-\frac{\partial^2 \log p}{\partial \theta_a^C\partial r_i^t} \Delta r_i
\end{equation}
Computing the expectation with respect to $n_i$ at each bin and then summing over bins, we arrive at 
\begin{eqnarray}
    &&\Delta \theta_b^C \sum_i\sum_{n_i}p(n_i|\vect{\theta}) \frac{\partial^2 \log p}{\partial \theta^C_a \partial \theta^C_b} \nonumber \\
    = &&-\sum_i\Delta r_i \sum_{n_i}p(n_i|\vect{\theta}) \frac{\partial^2 \log p}{\partial \theta_a^C\partial r_i^t}.
\end{eqnarray}
If we further notice
\begin{equation}
    \sum_{n_i} p(n_i|\vect{\theta}) \frac{\partial^2 \log p}{\partial \theta_a\theta_b} 
    = -\sum_{n_i} p(n_i|\vect{\theta}) \frac{\partial \log p}{\partial \theta_a}\frac{\partial \log p}{\partial \theta_b}, 
\end{equation}
we arrive at
\begin{equation}
    \Delta \vect{\theta}^C = -\left[\vect{I}^C\right]^{-1}
    \int 
    \left[\frac{\partial \log r(x|\vect{\theta})}{\partial \vect{\theta}^C} \right]
    \Delta r(x) dx.
    \label{eq:bias_dr_on_thetac}
\end{equation}
We can thus use Eq.~(\ref{eq:bias_dr_on_thetac}) to study how an error in the astrophysical rate model, $\Delta r(x)$,  propagates to the cosmological parameters, $\vect{\theta}^C$. Note that while we focus on $\Delta \vect{\theta}^C$ in this study, our framework can also be straightforwardly extended to study the bias on astrophysical parameters.

\section{Combined astrophysical and cosmological model}
\label{sec:model}

In this section, we derive the expected event rate $r(m_1, m_2, z | \vect{\theta})$ which can then be used to construct the Fisher information [Eq.~(\ref{eq:fisher})] and/or estimate the bias on $\vect{\theta}^C$ [Eq.~(\ref{eq:bias_dr_on_thetac})]. 

Suppose the intrinsic distribution of GW events is~\citep{Fishbach:18, LVK:21cos} 
\begin{equation}
    \frac{dn}{dm_1 dm_2 dz}(m_1, m_2, z |\vect{\theta}) = R~p(m_1, m_2, z|\vect{\theta}),
\end{equation}
where $R$ is total number of BBHs and we normalize the probabilities such that
\begin{equation}
    \int dm_1 dm_2 dz ~p(m_1, m_2, z|\vect{\theta}) = 1.
\end{equation}

The expectation of the observed event density is 
\begin{align}
    &r(m_1, m_2, z| \vect{\theta}) =  \frac{dn_{\rm obs}}{dm_1 dm_2 dz}(m_1, m_2, z| \vect{\theta})  \nonumber \\
    =& R P_{\rm det}[m_1, m_2, D_L(z | \vect{\theta}^C) ]  p(m_1, m_2, z|\vect{\theta}),
    \label{eq:r_m1m2z}
\end{align}
where $D_L$ is the luminosity distance and $P_{\rm det}\in [0, 1]$ is the fraction of GW events with $(m_1, m_2, z)$ that are detectable. 

The above expression is generic. To proceed, we further make simplifying assumptions following \citet{Fishbach:18} and consistent with \citet{LVK:21cos}. In particular, we assume
\begin{equation}
    p(m_1, m_2, z|\vect{\theta}) = p(m_1, m_2 | \vect{\theta}^A)p(z | \vect{\theta}^A, \vect{\theta}^C), 
\end{equation}
where $p(m_1, m_2 | \vect{\theta}^A)$ describes the mass distribution and we assume that it is independent of the redshift. The redshift distribution is then captured by  $p(z | \vect{\theta}^A, \vect{\theta}^C)$. We separately normalize the two distributions as $\int p(m_1, m_2 | \vect{\theta}^A) dm_1 dm_2 {=} 1$ and $\int p(z | \vect{\theta}^A, \vect{\theta}^C) dz{=}1$. %\bs{One interesting bias that could be tested is whether the mass distribution evolves with redshift due to eg upper mass gap evolving with redshift because the metallicity changes. }

For the rest of our study, we will focus on the case where $p(m_1, m_2 | \vect{\theta}^A)$ is described by the \texttt{Power Law $+$ Peak} model~\citep{Talbot:18, LVC:21gwtc2pop} and we use the same notation as used in \citet{LVK:21cos}. In this case, the distribution of the mass of the primary BH, $m_1$, (with $m_1\geq m_2$) contains two components: a truncated power-law component defined between $(M_{\rm min}, M_{\rm max})$ with $p(m_1)\propto m_1^{-\alpha}$, and a Gaussian peak centered at $\mu_g$ and with a width of $\sigma_g$. The overall height of the Gaussian peak is governed by a parameter $\lambda_g$. For a given $m_1$, the secondary mass then follows a truncated power-law between $(M_{\rm min}, m_1)$  with a slope $p(m_2)\propto m_2^{\beta}$. Additionally, we smooth the lower end of both $m_1$ and $m_2$ with a sigmoid function defined in eq. (B7) in \citet{LVK:21pop} and with a parameter $\delta_m$. %In summary, the mass distribution is described by 8 parameters in total, including $\left(M_{\rm min}, M_{\rm max}, \delta_m, \alpha, \beta, \lambda_g, \mu_g, \sigma_g\right)$ and our notation follows \citet{LVK:21cos}. 

For the redshift model, we further write
\begin{equation}
    p(z | \vect{\theta}^A, \vect{\theta}^C) \propto \frac{dV_c}{dz}\left(z|\vect{\theta}^C\right) \frac{\psi(z|\vect{\theta}^A)}{1+z},
    \label{eq:p_z}
\end{equation}
where $V_c(z|\vect{\theta}^C)$ is the comoving volume and the $1/(1+z)$ term converts from detector-frame to source-frame time. A general parameterization of the $\psi(z)$ piece can be written as~\citep{Madau:14} 
\begin{equation}
    \psi(z) = \left[1+(1+z_p)^{-\gamma-k}\right]\frac{(1+z)^\gamma}{1+\left[(1+z)/(1+z_p)\right]^{\gamma+k}}, 
\end{equation}
where $\gamma$ and $k$ respectively describe the low- and high-redshift power-law slopes and $z_p$ corresponds to a peak in $\psi(z)$. For GWTC-3 where most events are detected at low redshifts, $\psi(z)$ simplifies to (see, e.g., \citealt{Fishbach:18})
\begin{equation}
    \psi(z) = \left(1+z\right)^\gamma. 
    \label{eq:psi_low_z}
\end{equation}
We will adopt Eq.~(\ref{eq:psi_low_z}) for our analysis and drop $(z_p, k)$.

Under the model described above, there are 9 astrophysical parameters $\vect{\theta}^A=\left(M_{\rm min}, M_{\rm max}, \delta_m, \alpha, \beta, \lambda_g, \mu_g, \sigma_g, \gamma\right)^{\rm T}$. For the cosmological part, we assumed a flat universe described by $\vect{\theta}^C = (H_0, \Omega_m)^{\rm T}$ with $H_0$ the Hubble constant and $\Omega_m$ the mass density normalized by the critical density. For future convenience, we will  define $h=H_0/(100\,{\rm km\,s^{-1}\,Mpc^{-1}})$. 

To estimate $P_{\rm det}$, we follow \citet{Fishbach:18} and approximate the observed signal-to-noise ratio (SNR) of an event as
\begin{equation}
    \rho[m_1, m_2, D_L(z | \vect{\theta}^C)] = \rho_0 \Theta,
    \label{eq:rho_approx}
\end{equation}
where $\rho_0$ is a characteristic SNR of the source and $\Theta$ accounts for the change in the SNR due to angular projection, with 
\begin{align}
    &\log \Theta \sim \mathcal{N}\left(0, \sigma^2_{\log\Theta} \right), \\
    &\sigma^2_{\log\Theta} = \frac{\sigma^2_{\log\Theta,0}}{1+{\rho_0}/{\rho_{\rm th}}}
    \label{eq:sigma_sq_logTh}
\end{align}
where $\sigma^2_{\log\Theta,0}$ and $\rho_{\rm th}$ are further parameters controlling the shape of $\Theta$. %\ali{rephrase: follows a log-normal distribution with mean.. and variance...?}

Suppose sources with $\rho > \rho_{\rm th}$ are detectable, we have
\begin{align}
    P_{\rm det} & = \int_{\log \Theta_{\rm th}}^{\infty}p(\log \Theta) d\log\Theta \nonumber \\
    & = \frac{1}{2}{\rm Erfc}\left(\frac{\log\Theta_{\rm th}}{\sqrt{2}\sigma_{\log\Theta}}\right).
    \label{eq:P_det}
    % P_{\rm det} &= \frac{\int_{\log \Theta_8}^\infty p(\log \Theta) \exp(\log \Theta) d\log \Theta}{\int_{-\infty}^\infty p(\log \Theta) \exp(\log \Theta) d\log \Theta} \nonumber \\
    % &=\frac{1}{2}\left(1 + {\rm Erf}\left[\frac{\sigma_\Theta^2 - \log \Theta_{8}}{\sqrt{2}\sigma_\Theta}\right) \right],
\end{align}
where $\Theta_{\rm th} = \rho_{\rm th} / \rho_0$ and ${\rm Erfc}$ is the complementary error function.

% Following \citet{Fishbach:18} and consistent with \citet{LVK:21cos}, we further assume
% \begin{equation}
%     p(m_1, m_2, z|\vect{\theta}) = p(m_1, m_2 | \vect{\theta}^A)p(z | \vect{\theta}^A, \vect{\theta}^C), 
% \end{equation}
% where $p(m_1, m_2 | \vect{\theta}^A)$ describes the mass distribution and we assume that it is independent of the redshift. For the redshift model, we further write
% \begin{equation}
%     p(z | \vect{\theta}^A, \vect{\theta}^C) = \frac{dV_c}{dz}\left(z|\vect{\theta}^C\right) \frac{\psi(z|\vect{\theta}^A)}{1+z},
% \end{equation}
% where $V_c(z|\vect{\theta}^C)$ is the comoving volume and the $1/(1+z)$ term converts from detector-frame to source-frame time.

\section{Applications to GWTC-3}
\label{sec:GWTC3}

\begin{table*}[!tb]
\centering
\caption{Values of $(\vect{\theta}^A, \vect{\theta}^C)$ used in our study to construct the Fisher information matrix (Eq.~(\ref{eq:fisher})) and estimate the bias due to $\Delta r$ (Eq.~(\ref{eq:bias_dr_on_thetac})). 
}
\begin{tabular}{ ccccccccccc } 
 $M_{\rm min}$ & $M_{\rm max}$ & $\delta_m$ & $\alpha$ & $\beta$ & $\lambda_g$ & $\mu_g$ & $\sigma_g$ & $\gamma$  & $h$ & $\Omega_m$\\
 \hline
 $6.5\, M_\odot$ & $112.5\,M_\odot$ & $2.5\,M_\odot$ & $3.78$ & $-0.81$ & $0.03$ & $32.27\,M_\odot$ & $3.88\,M_\odot$ & $4.59$ &  $0.7$ & $0.3$
\end{tabular}
\label{tab:pars}
\end{table*}

In this section, we apply our method to GWTC-3~\citep{LVK:21gwtc3} and estimate the uncertainties on $(\vect{\theta}^A, \vect{\theta}^C)$ when jointly fitting the astrophysical population distribution and cosmology together. Despite the simplicity of our method, it successfully captures many qualitative features and gives accurate predictions on different parameters' uncertainties as reported in \citet{LVK:21cos}. Further validation of our method can be found in Appx.~\ref{sec:validation} where we also apply our method to reproduce results in \cite{Fishbach:18, Farr:19} .  

Note that to evaluate the Fisher information matrix (Eq.~(\ref{eq:fisher})), we need to take derivatives around the ``true'' model parameters. These values are mostly approximated by the ones inferred in \citet{LVK:21cos} and we summarize them in Table~\ref{tab:pars}. Figs.~\ref{fig:sample_dist_vs_z} and \ref{fig:sample_dist_vs_m1det} are also generated with the same set of parameters (except for the one listed in the legend).  
Note that we slightly modified the values of $M_{\rm min}=6.5\,{M_\odot}$ and $\delta_m = 2.5\,{M_\odot}$ to make our Fig.~\ref{fig:sample_dist_vs_m1det} more similar to fig. 1 in \citet{LVK:21cos}.\footnote{There are likely two peaks in the mass distribution as suggested in \citet{LVK:21pop} and the lower one (around $m_1=10\,M_\odot$) is not captured by the \texttt{Power Law + Peak} model adopted by \citet{LVK:21cos}.}
The overall scale $R$ is set so that the total number of BBH detection is $n_{\rm obs}=\int r(m_1, m_2, z|\vect{\theta}) dm_1dm_2dz=40$, consistent with the number of BBH events used in \cite{LVK:21cos}. 

To approximate $P_{\rm det}$, we compute the characteristic $\rho_0$ using a single detector with LIGO Hanford's sensitivity in the third observing run~\citep{Buikema:20} for each $[m_1, m_2, D_L(z|\vect{\theta}^C)]$. The waveform is generated with the IMRPhenomD approximation (\citealt{Khan:16}; the waveform is computed using \texttt{PYCBC}~\citealt{Nitz:22}) and the source is placed at an effective distance of $2.3 D_L$~\citep{Allen:12}. 
We further use $\rho_{\rm th}=8$ and $\sigma_{\log \Theta, 0}^2=0.25$ when computing Eq.~(\ref{eq:P_det}).

\subsection{Using redshift distribution while holding population model fixed}

Firstly, we consider the case where we constrain the cosmological parameters using the redshift distribution of BBH events while treating the underlying astrophysical population as known and fixed. An astrophysical expectation  can be constructed using the coarse-grained distribution of galaxies. Indeed, when each BBH event is localized with limited accuracy and thousands of galaxies or more lie within the uncertainty volume, a galaxy catalog mainly serves as an estimation of the overall, smoothed shape of $\psi(z)$, which we model as a simple power law as in Eq.~(\ref{eq:psi_low_z}). In this case, cosmological parameters are constrained by requesting consistency between the distribution of observed BBH events and our astrophysical expectation, as demonstrated in the upper panel of  Fig.~\ref{fig:sample_dist_vs_z}. (We will return to this later in Sec.~\ref{sec:conclusion_discussion} to discuss how an improved localization accuracy together with a complete galaxy catalog could help.) 

\begin{figure}
  \centering
  \includegraphics[width=\columnwidth]{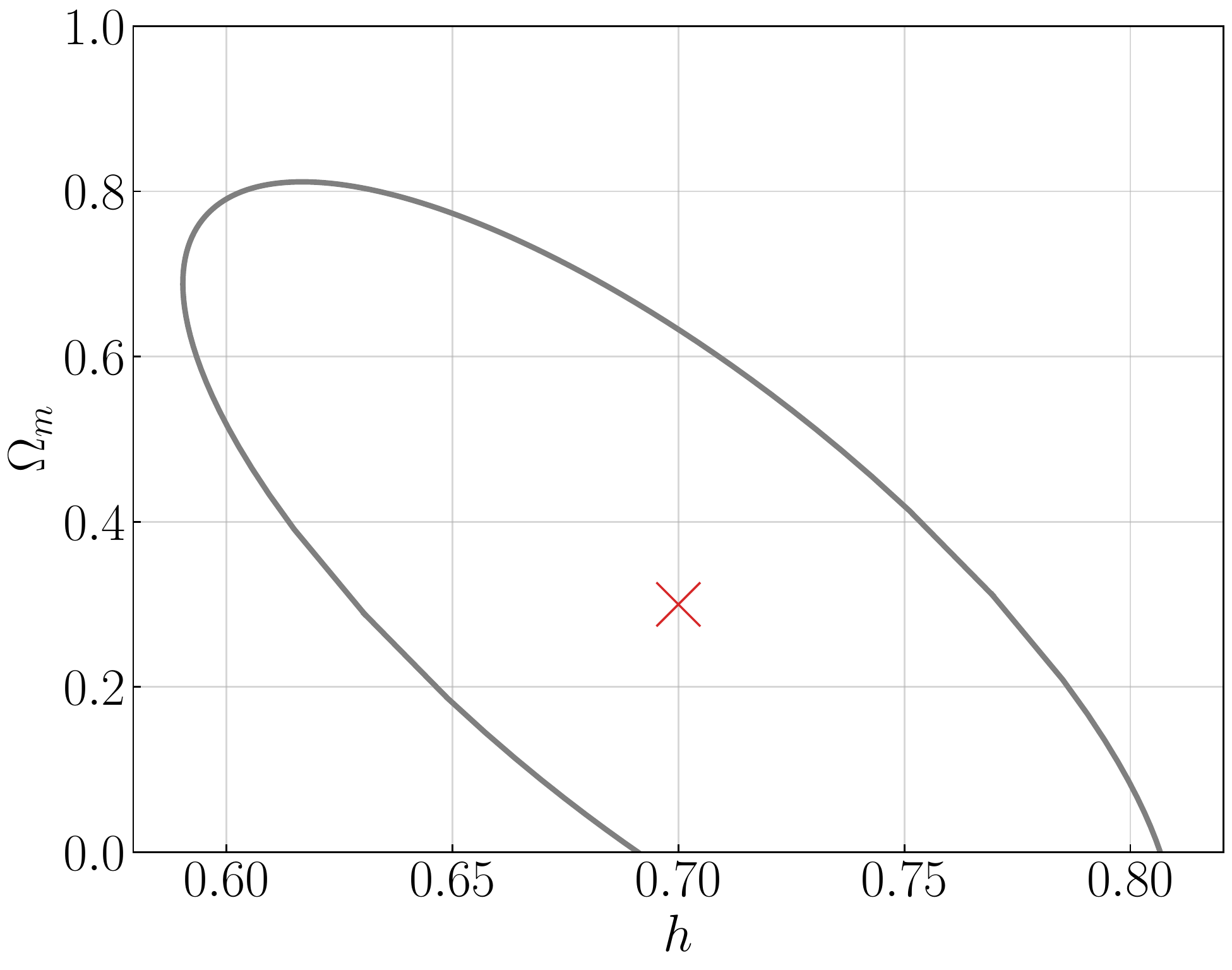}
  \caption{Uncertainties on cosmological parameters $(h, \Omega_m)$ from the redshift histogram (cf. Fig.~\ref{fig:sample_dist_vs_z}) assuming we know exactly the astrophysical model. 
  Throughout this work, we will use red crosses to denote the true values of the parameters (i.e., values at which we evaluate the Fisher information matrix). The error ellipses indicate the $68\%$ credible intervals.
  We predict an uncertainty on $h$ of $\pm0.11$, which agrees well with the gray-dotted curve in fig. 9 in \citet{LVK:21cos} obtained under the same assumptions.
  %It agrees nicely with fig. 9 in \cite{LVK:21cos}.  
  }
\label{fig:GWTC3_dh_fix_pop}
\end{figure}

In Fig.~\ref{fig:GWTC3_dh_fix_pop}, we present the constraints on $(h, \Omega_m)$ from the marginalized redshift distribution $r(z|\vect{\theta}){=}\int r(m_1, m_2, z|\vect{\theta})dm_1 dm_2$ (cf. Fig.~\ref{fig:sample_dist_vs_z}). The result is obtained by inverting a $3\times3$ Fisher matrix involving $(h, \Omega_m, R)$ and treating $\vect{\theta}^A$ as known (Eq.~(\ref{eq:fisher}) with $x$ replaced by $z$). Our approach predicts an uncertainty in $h$ to be 0.11, nicely agreeing with the results shown in Fig. 9 in \citet{LVK:21cos}. $\Omega_m$, on the other hand, is not well constrained (in fact, its error is greater than its true value and thus it exceeds the capability of Fisher matrix) because of both the relatively small sample size ($n_{\rm obs}=40$) and the fact that most events are detected at low redshift with $z < 0.5$. 

\begin{figure}
  \centering
  \includegraphics[width=\columnwidth]{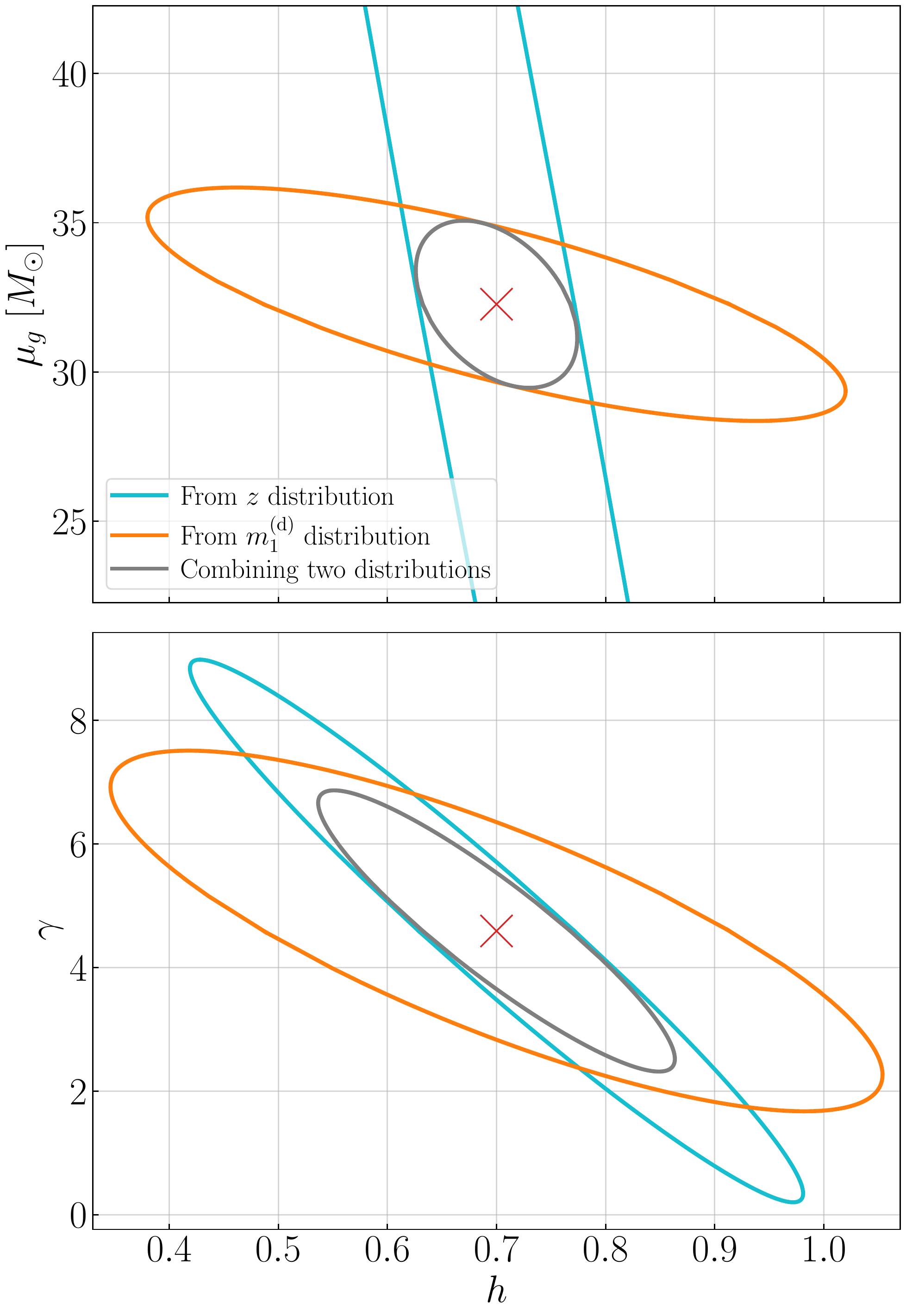}
  \caption{Correlation between astrophysical and cosmological parameters by inverting a $3\times3$ Fisher matrix including $(h, \mu_g, R)$ (top panel) or $(h, \gamma, R)$ (bottom panel). The cyan ellipses correspond to constraints from the redshift distribution alone (cf. Fig.~\ref{fig:sample_dist_vs_z}). As $\mu_g$  and/or $\gamma$  decreases, $h$ will increase to a greater value. It captures the key features shown in fig. 10 in \cite{LVK:21cos}.  If one further incorporates the information from the mass distribution (orange ellipses; cf. Fig.~\ref{fig:sample_dist_vs_m1det}), the combined uncertainties can be reduced to the gray ellipses. 
  }
\label{fig:GWTC3_dh_from_z_dist}
\end{figure}

However, as pointed out in, e.g., \citet{Mastrogiovanni:21, LVK:21cos}, and illustrated in Fig.~\ref{fig:sample_dist_vs_z}, the constraints on the cosmological parameters rely critically on the assumptions of the astrophysical model. We  elaborate on this point further in Fig.~\ref{fig:GWTC3_dh_from_z_dist} in the cyan error ellipses. 
We obtained these ellipses by inverting a $3\times3$ Fisher matrix involving $(h, \mu_g, R)$ in the top panel and one involving $(h, \gamma, R)$ in the bottom panel. We notice strong anti-correlations between $h$ and $\mu_g$ and between $h$ and $\gamma$, consistent with the results shown in \citet{LVK:21cos}. This demonstrates that with the redshift distribution of BBH events alone, measuring cosmological parameters can be challenging 
unless we have a highly precise knowledge of the intrinsic population model. 
% \bs{How would this highly precise knowledge of the pop model ever be found, except by jointly fitting astro and cosmo? }
% \hang{(HY: This is basically impossible I would say. But we cannot say explicitly that LIGO's method is flawed.)}

\subsection{Jointly fitting astrophysical population model and cosmology}

Fortunately, besides the redshift distribution itself, we also have information on other properties of BBH events such as the mass distribution.
As demonstrated in Fig.~\ref{fig:sample_dist_vs_m1det}, the partial degeneracy between $h$ and $\mu_g$ shown in redshift distribution (Fig.~\ref{fig:sample_dist_vs_z}) can be largely broken once we include the distribution of the detector-frame mass distribution of the primary, $r[m_1^{(d)}|\vect{\theta}]=\int\left[r(m_1, m_2, z)/(1+z)\right]dm_2 dz$. 

Similar to how we obtain the cyan ellipses in Fig.~\ref{fig:GWTC3_dh_from_z_dist}, we also construct Fisher matrices for $(h, \mu_g, R)$ in the top panel (or $(h, \gamma, R)$ in the bottom panel) from the $m_1^{\rm (d)}$ distribution. The results are shown in the orange ellipses. Since distributions of both  $z$ and $m_1^{\rm (d)}$ are available in a GW catalog, we can combine them together, leading to the gray ellipses in Fig.~\ref{fig:GWTC3_dh_from_z_dist}. This allows us to individually constrain $h$ and $\mu_g$ to good accuracy (assuming other parameters in $\vect{\theta}$ are known), and the covariance between $h$ and $\gamma$ can also be significantly reduced.

\begin{figure*}
  \centering
  \includegraphics[width=1.9\columnwidth]{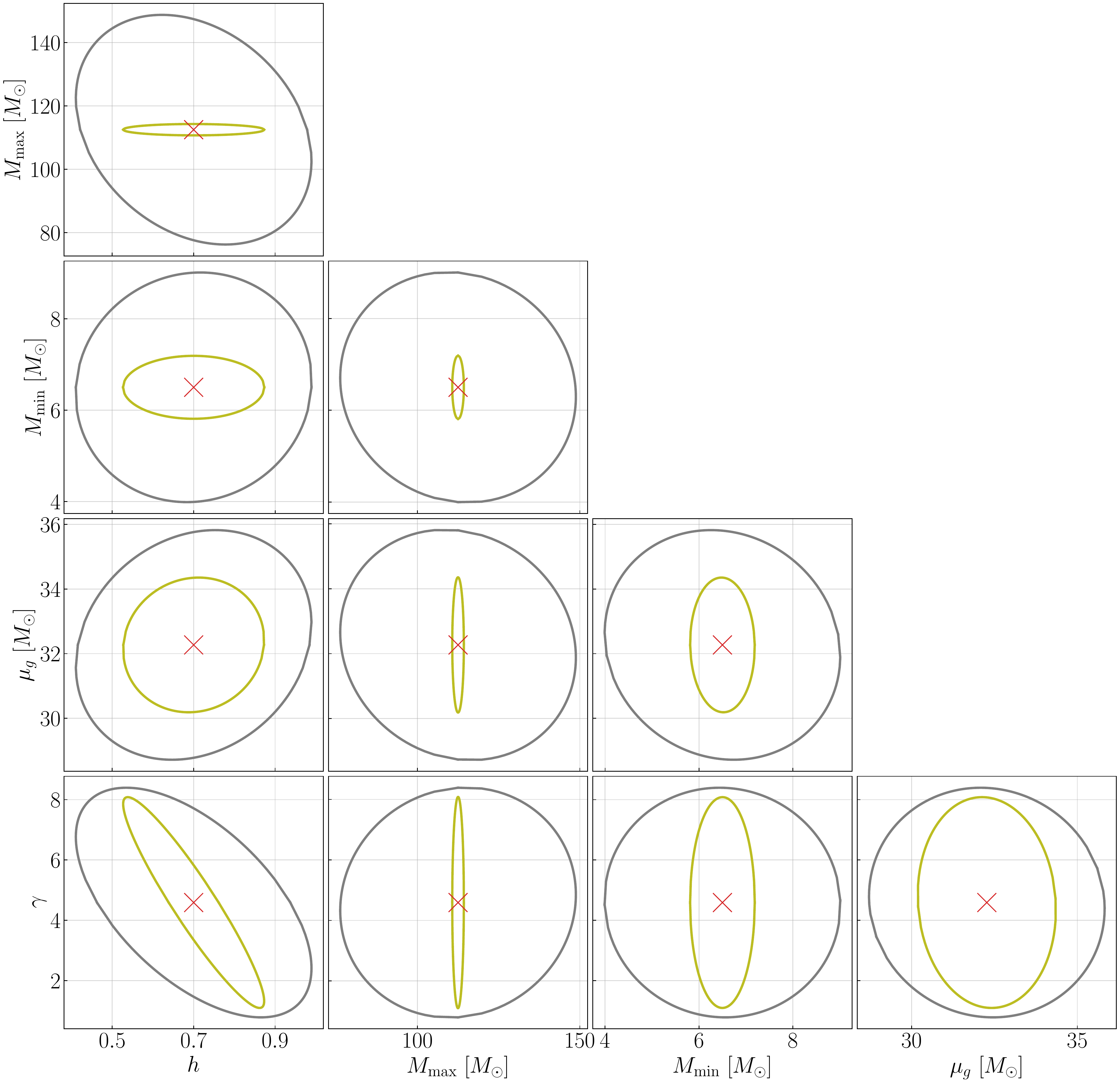}
  \caption{Error ellipses for a sample of 40 BBH events similar to the GWTC3 catalog. 
  The gray ellipses are obtained by summing the Fisher information from the marginalized redshift and primary mass distribution together, and the olive ones are from the 3D $r(m_1, m_2, z)$ distribution. 
  %Due to the relative small sample size, the gray ellipses show better agreement with the results reported in \citet{LVK:21cos} than the olive ones. With more events to fill in the 3D distribution, the result may change from the gray ellipses to the olive ellipses (in addition to the $1/\sqrt{n_{\rm obs}}$ scaling). 
%   The uncertainty on each quantity matches the results presented in \cite{LVK:21cos}. The direction of correlation is different for $\mu_g$ and $h$ because we only use a toy model for $P_{\rm det}$. 
  }
\label{fig:GWTC3_fisher_err}
\end{figure*}

Combining the Fisher information from the redshift and mass distributions together is largely similar to the hierarchical inference performed in \citet{LVK:21cos}. To illustrate this point, we now invert the full Fisher matrix (note that in Fig.~\ref{fig:GWTC3_dh_from_z_dist} we considered only submatrices) and the results are shown in Fig.~\ref{fig:GWTC3_fisher_err}. More specifically, we construct two Fisher matrices using Eq.~(\ref{eq:fisher}) with $x$ respectively substituted by $z$ and $m_1^{(d)}$. The two matrices are summed together and then inverted to give us the gray error ellipses. 

Overall, our result shows nice agreement with the one reported in \citet{LVK:21cos}. In particular, the $68\%$ credible interval for $h$ is $h=0.70\pm0.29$ and it exhibits a strong anti-correlation with $\gamma$ and $M_{\rm max}$ whose uncertainties are also consistent with fig.~5 in \citet{LVK:21cos}. Because we used a simple approximation of $P_{\rm det}$ [Eqs.~(\ref{eq:rho_approx})-(\ref{eq:P_det})] and we ignored the statistical error on each individual event, we do not expect an exact reproduction of the results in \citet{LVK:21cos}. Due to our simplifying treatments, $\mu_g$ is better constrained than in \citet{LVK:21cos} and its correlation with $h$ as well as with other parameters is lifted (see also Fig.~\ref{fig:GWTC3_dh_from_z_dist} and note the gray error ellipse in the upper panel is much smaller than the one in the bottom panel).

In fact, we can directly construct a Fisher matrix from a 3-dimensional (3D) distribution $r(m_1, m_2, z|\vect{\theta})$. This leads to the olive ellipses in Fig.~\ref{fig:GWTC3_fisher_err}. This contains more information and thus leads to tighter constraints on parameters compared to combining two marginalized distributions (gray ellipses). 
For GWTC-3 with only slightly more than 40 BBH events, however, we do not have a high ``SNR'' in the 3D histogram $r(m_1, m_2, z|\vect{\theta})$.\footnote{
% Consider a discrete example where we divide $m_1$ into $n_m\approx (M_{\rm max}-M_{\rm min})/\sigma_g\simeq 25$ bins so that we can resolve the Gaussian peak(s) in the mass histogram. The 2D histogram for $(m_1, m_2)$ would thus need $\approx n_m(n_m-1)/2=300$ bins. Even we just construct a coarse-grained redshift distribution with a few bins, a size of $\mathcal{O}(1,000)$ bins is likely needed for the 3D distribution. This would require at least $\mathcal{O}(1,000)$ events for each bin to contain one event. This is much greater than the sample size of GWTC-3 yet achievable when aLIGO reaches its designed sensitivity (as assumed in, e.g., \citealt{Farr:19}).  
Consider a discrete example. We would need at least 8 different bins to constrain $\left(M_{\rm min}, M_{\rm max}, \delta_m, \alpha, \lambda_g, \mu_g, \sigma_g\right)$ in the histogram of $m_1$ or $m_1^{\rm (d)}$. For the secondary mass $m_2$, we would additionally need 2 more bins to determine the power-law slope $\beta$. The redshift distribution requires at least 3 bins to constrain $(\gamma, h)$. Thus a full 3D histogram would require more than 48 bins. This is greater than the sample size used by ~\citet{LVK:21cos}. Nonetheless, there will be enough events to populate the 3D histogram when aLIGO reaches its designed sensitivity and detects $\mathcal{O}(1,000)$ events per year (as assumed in, e.g., \citealt{Farr:19}). 
\label{ft:3dSNR}
}
Therefore, summing marginalized distribution in $z$ and in $m_1^{(d)}$ (gray ellipses) provides a better agreement of GWTC-3 results~\citep{LVK:21cos} than the 3D distribution (olive ellipses). Nonetheless, as the sample size increases, we would expect that the 3D distribution becomes a more accurate prediction (which we validate in Appendix~\ref{sec:validation} by reproducing the results in \citealt{Fishbach:18, Farr:19}). Therefore, in addition to the $1/\sqrt{n}_{\rm obs}$ reduction in the uncertainties (as obviously seen in Eqs.~(\ref{eq:fisher}) and (\ref{eq:r_m1m2z})), we would expect the results reported in \citet{LVK:21cos} to improve further from the gray ellipses to the olive ones as the SNR of each bin in the 3D distribution increases (with the expectation of the bin becomes greater than its Poissonian error; see Footnote~\ref{ft:3dSNR}). This can be especially valuable for constraining $M_{\rm max}$ as changing it can significantly alter $P_{\rm det}$ at large redshift, a point we will illustrate further when discussing the bias on cosmological parameters. 

\section{Bias induced by substructures in the population model}
\label{sec:bias}

Having discussed in the previous section the parameter estimation uncertainties when jointly fitting the cosmological and astrophysical models, we now consider the bias in the cosmological parameters (especially $H_0$) induced by inaccuracies in our astrophysical model, which is naturally expected if our parameterized model is insufficient to capture all the details in the true population model. Indeed, we note that the specific functional form assumed in our study (the \texttt{Power Law+ Peak} model) is not significantly preferred over, e.g., a \texttt{Broken Power Law} model~\citep{LVK:21cos}. More possibilities with different parametrizations are also considered in, e.g., \citet{LVK:21pop, Roulet:21}. Furthermore, the mass distribution could contain more complicated features~\citep{Tiwari:21} and/or be redshift dependent~\citep{Mukherjee:21b, Mapelli:22,vanSon:22, Karathanasis:22}, introducing more features beyond what is captured by the model described in Sec.~\ref{sec:model}. Similarly, an error in the redshift model $\psi(z)$ could also bias the inferred cosmology~\citep{You:21}.

% Such an error can happen due to e.g. substructures that are not captured by the parameterized model $p(m_1, m_2, z|\vect{\theta})$.

Suppose the true event density  can be written as 
\begin{align}
    &r^{t}(m_1, m_2, z)=RP_{\rm det} \nonumber \\
    \times&  \left[(1-\Delta r_0)p(m_1, m_2, z|\vect{\theta}) +\Delta r_0 p_{\rm err}(m_1, m_2, z)\right],
\end{align}
and our parameterized model captures the $r{=}RP_{\rm det} p(m_1, m_2, z|\vect{\theta})$ part.
This leads to an error of
\begin{equation}
    \Delta r(m_1, m_2, z) = \Delta r_0 \left[RP_{\rm det} p_{\rm err}(m_1, m_2, z) - r(m_1, m_2, z)\right],
    \label{eq:delta_r}
\end{equation}
where $p_{\rm err}$ specifies the shape of the deviation and it is normalized to $\int  p_{\rm err}(m_1, m_2, z)dm_1 dm_2 dz=1$, and $\Delta r_0$ is an overall factor governing the magnitude of the deviation. We note further that the $-r$ term only affects the overall number of GW events when plugged into Eq.~(\ref{eq:bias_dr_on_thetac}) and therefore can be absorbed by a rescaling of $R$; when $\Delta r_0>0$, it decreases the value of $R$. For the rest of the section we will focus on the effect induced by $p_{\rm err}$. 

In particular, we focus on bias induced by unmodeled local substructures. For this, we write 
\begin{equation}
    p_{\rm err}(m_1, m_2, z) = p_{\rm err}(m_1, m_2) p_{\rm err}(z),
    \label{eq:p_err}
\end{equation}
with
\begin{align}
    &p_{\rm err}(m_1, m_2) \propto \frac{1}{m_1 - M_{\rm min}} 
    \exp\left[\frac{-(m_1 - \mu_{m,{\rm err}})^2}{2\sigma^2_{m, {\rm err}}}\right], 
    \label{eq:p_err_m}\\
    &p_{\rm err}(z) \propto \frac{1}{1+z}\frac{dV_c}{dz}\exp\left[\frac{-(z - \mu_{z,{\rm err}})^2}{2\sigma^2_{z, {\rm err}}}\right],
    \label{eq:p_err_z}
\end{align}
where the location of the substructure is governed by $\mu_{m,{\rm err}}$ and $\mu_{z,{\rm err}}$ and width by $\sigma_{m, {\rm err}}$ and $\sigma_{z, {\rm err}}$. In our study, we vary $\left(\mu_{m,{\rm err}}, \mu_{z,{\rm err}}\right)$ and fix $\sigma_{m, {\rm err}}=1\,M_\odot$ and $\sigma_{z, {\rm err}}=0.025$. As a brief aside, we note that the local error considered here can serve as the building block for considering more extended errors, as a generic $\Delta r$ can be viewed as the superposition of many such local substructures. 

To set the overall factor $\Delta r_0$, we request
\begin{equation}
    \frac{\Delta r_0 \int P_{\rm det}p_{\rm err}dm_1dm_2dz}{\int P_{\rm det}dm_1dm_2dz}=0.01. 
\end{equation}
In other words, we assume the unmodeled substructure contains $1\%$ of the BBH events. Note that we choose $\Delta r_0>0$ for the simplicity of our discussion, $\Delta r_0$ can be either positive (a local peak) or negative (a local trough).

In this section, we follow \citet{Fishbach:18} and approximate $P_{\rm det}$ according to the aLIGO design sensitivity. In particular, we approximate the characteristic SNR as 
\begin{equation}
    \rho_0 = 8 \left[\frac{\mathcal{M}_c(1+z)}{10\,{M_\odot}}\right]^{5/6}\left(\frac{1\,{\rm Gpc}}{D_L}\right),
    \label{eq:rho_char_FHF18}
\end{equation}
where $\mathcal{M}_c=m_1^{3/5}m_2^{3/5}/(m_1+m_2)^{1/5}$ is the chirp mass of the BBH. Following~\citet{Fishbach:18}, we further set $\rho_{\rm th}=8$ and $\sigma_{\log\Theta,0}^2=0.3$ in Eq.~(\ref{eq:sigma_sq_logTh}).

\begin{figure}
  \centering
  \includegraphics[width=\columnwidth]{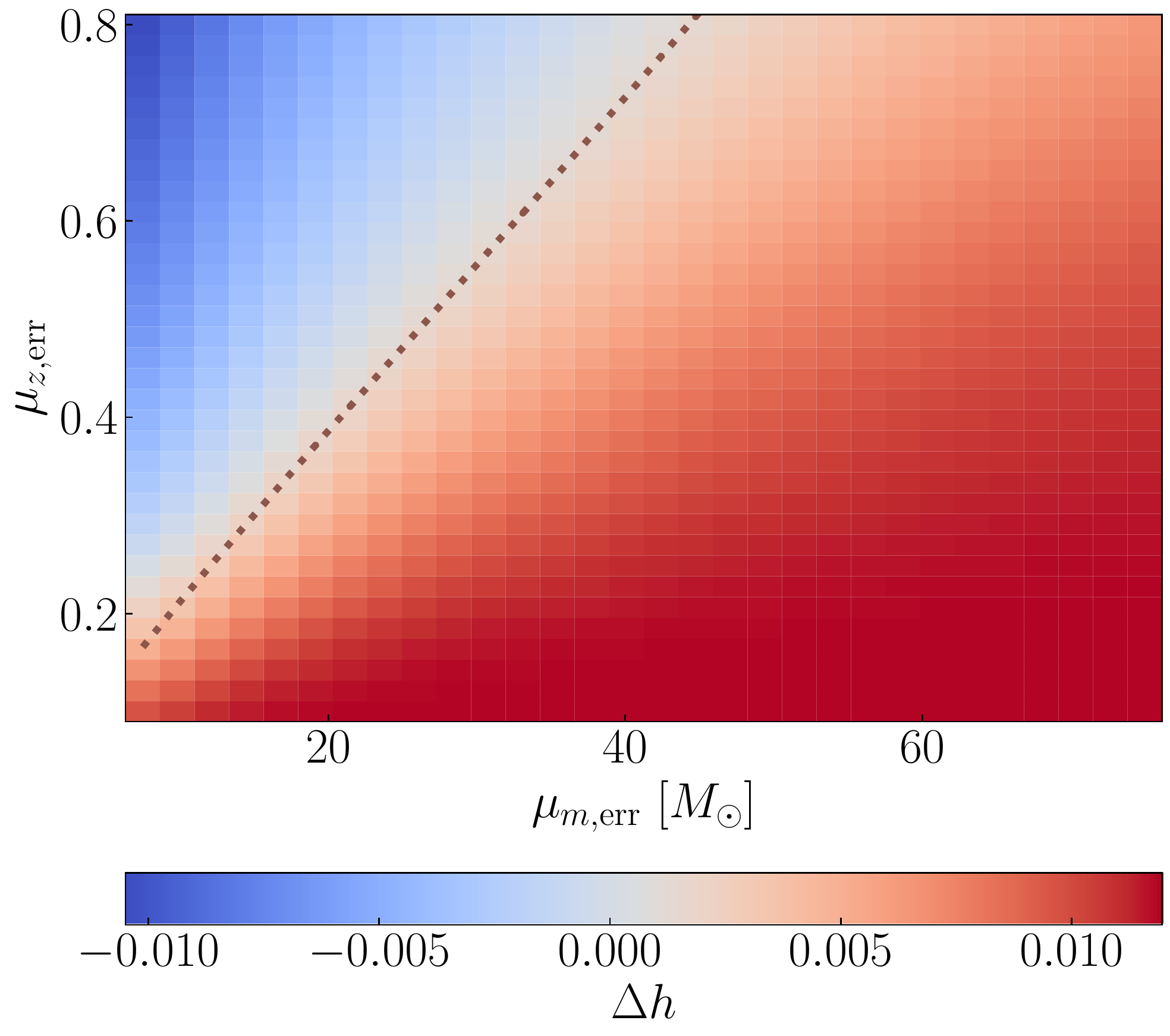}
  \caption{Bias on $h$ due to an error in the astrophysical rate $\Delta r$ given by Eqs.~(\ref{eq:p_err})-(\ref{eq:p_err_z}). An error in $m_1$ but constant in redshift can then be obtained by summing over all the pixels along a specific $\mu_{m,{\rm err}}$ (i.e., a vertical line) with appropriate normalization. Likewise, other generic $\Delta r$ can be obtained by summing over the corresponding pixels. Also shown in the dotted-brown line is an approximation of the detection threshold with $\left[\mathcal{M}_c(1+z)\right]^{5/6} / z\simeq {\rm constant}$. 
  }
\label{fig:bias_on_h}
\end{figure}

We are now ready to evaluate the bias due to $\Delta r$ (Eq.~(\ref{eq:delta_r})) on cosmological parameters according to Eq.~(\ref{eq:bias_dr_on_thetac}). Here we focus on the bias on $h$ and we consider $\vect{\theta}^C{=}(h, R)^{\rm T}$ in Eq.~(\ref{eq:bias_dr_on_thetac}). The result is shown in Fig.~\ref{fig:bias_on_h}. 

Firstly, we note that the bias is independent of $n_{\rm obs}$. This is because in Eq.~(\ref{eq:bias_dr_on_thetac}) we have $\left[\vect{I}^C\right]^{-1}\propto n_{\rm obs}^{-1}$ whereas $\Delta r\propto n_{\rm obs}$. This is in contrast to the statistical uncertainty discussed in Sec.~\ref{sec:GWTC3} which reduces as $n_{\rm obs}^{-1/2}$. Therefore, while we expect a significant reduction in the statistical uncertainty as current detects become increasingly more sensitive, and as the 3G GW detectors like Cosmic Explorer \citep{Reitze:19} and Einstein Telescope \citep{Sathyaprakash:11} come online in 2030s, the systematic bias would persist unless we incorporate more sophisticated models. In particular, we would expect to detect 15,000 BBH events every month with 3G detector~\citep{Vitale:19}. This means we would reduce the statistical error on $h$ to sub-percent level within a month of observation according to Fig.~\ref{fig:GWTC3_fisher_err}. This is below the bias shown in Fig.~\ref{fig:bias_on_h} and therefore the dark siren cosmology would be limited by uncertainties in our astrophysical population model. 

We further note that for large $\mu_{m,{\rm err}}$ and small $\mu_{z,{\rm err}}$ (the bottom-right part of Fig.~\ref{fig:bias_on_h}), the bias is nearly a constant. 
The bias then gradually decreases and then becomes negative as $\mu_{m,{\rm err}}$ decreases and the $\mu_{z,{\rm err}}$ increases, or as we go to the top left part of Fig.~\ref{fig:bias_on_h}. The transition is characterized by the line of $\rho_0=8$ (the brown-dotted line in Fig.~\ref{fig:bias_on_h}), where we have used $m_1=m_2=\mu_{m,{\rm err}}$ to evaluate $\mathcal{M}_c$ and $\mu_{z,{\rm err}}$ to evaluate $D_L$ in Eq.~(\ref{eq:rho_char_FHF18}). 

These features can be understood as the following. 
Because we assume $p_{\rm err}$ is caused by local substructures and model it as a multivariate Gaussian in $m_1$ and $z$ (and uniform in $m_2$), from Eq.~(\ref{eq:bias_dr_on_thetac}) the bias is approximately given by\footnote{Here we treat $p(z)$ as an un-normalized function and use $R$ to absorb the normalization to simplify the discussion. Note that $h$ and $R$ are not completely degenerate because of $P_{\rm det}$, and it can be seen from Fig.~\ref{fig:sample_dist_vs_z}. In the real calculation, we include both $h$ and $R$ in $\vect{\theta}^{C}$ and hence $\vect{I}^C$ when evaluating Eq.~(\ref{eq:bias_dr_on_thetac}) to account for the correlation between them arising from this freedom in the definition of $p(z)$ and $R$. }
\begin{align}
    \Delta h 
    &\propto 
    \frac{\partial \log r(\mu_{m, {\rm err}}, \mu_{m, {\rm err}}, \mu_{z, {\rm err}}|\vect{\theta})}{\partial h}, \nonumber \\
    &\sim \frac{\partial \log P_{\rm det}}{\partial h} 
    + \frac{\partial \log \left( dV_c/dz\right)}{\partial h},
    \label{eq:bias_explanation}
\end{align}
where in the second line we have selected out the terms that have non vanishing derivatives with respect to $h$ and those values are approximately evaluated at $(m_1, m_2, z){=}(\mu_{m, {\rm err}}, \mu_{m, {\rm err}}, \mu_{z, {\rm err}})$. 

In the bottom-right part of Fig.~\ref{fig:bias_on_h}, $P_{\rm det}\simeq 1$. Thus the only contribution to $\Delta h$ comes from $\partial \log \left( dV_c/dz\right) / \partial h = 3/h$, which is a constant. This is why the bias is nearly constant in this region.  Physically, the excess events contained in $\Delta r$ makes us infer a greater comoving volume than the true value at a given redshift, which then leads to a positive bias in $h$.

As we move towards the top-left part of Fig.~\ref{fig:bias_on_h}, $P_{\rm det}$ changes from 1 to 0. Numerically, the slope is the steepest when $\Theta_{\rm th} = \rho_{\rm th}/\rho_0$ is around 1. Because changing $h$ changes the value of $\rho_0$ at a given redshift $\mu_{z, {\rm err}}$, the $\partial \log P_{\rm det}/\partial h$ term in Eq.~(\ref{eq:bias_explanation}) now starts to contribute. This drives changes the bias $\Delta h$ to a more negative value. Depending on the location, a local substructure containing $1\%$ of BBH events could bias the estimation of $h$ by about $1\%$ in either the positive or the negative direction. As we mentioned above, the statistical error on $h$ will drop below $1\%$ with about $10^4$ events. This is likely beyond aLIGO's expected detection number, yet it can be easily achieved with 3G detectors. Our study thus sets requirements of the accuracy of our astrophysical population model in the 3G era.

\section{Conclusion and Discussion}
\label{sec:conclusion_discussion}

In this study, we derived the Cram\'er-Rao bound of both astrophysical and cosmological parameters from the distributions (both marginalized and high-dimensional) of BBH events. Our approach complements the hierarchical inference currently employed by, e.g., \citet{LVK:21cos}. Its analytical simplicity makes it especially useful in predicting the performance of future detectors and providing insights in the statistics. 

The basic framework to both perform joint astrophysical and cosmological parameter estimations and compute bias in parameters due to errors in the assumed model was presented in Sec.~\ref{sec:framework}. The specific population model in our analysis was introduced in Sec.~\ref{sec:model}, which we then applied to place constraints on a BBH sample similar to GWTC3 in Sec.~\ref{sec:GWTC3}. In particular, we found that the GWTC3 results can be well reproduced if we combine the Fisher information of both the BBHs' redshift distribution and the mass distribution together. In the future, tighter constraints (in addition to the $\sqrt{n_{\rm obs}}$ reduction in the errors) would be expected as more events would allow us to construct an accurate 3D distribution of BBH events in the $(m_1, m_2, z)$ space. Then in Sec.~\ref{sec:bias}, we further considered the bias induced by unmodeled substructures in the population model. The bias due to other forms of $\Delta r$ can be readily obtained by summing over relevant pixels in Fig.~\ref{fig:bias_on_h} with proper reweighting. For instance, a substructure in $m_1$ but constant in $z$ can be obtained by summing along a vertical line in Fig.~\ref{fig:bias_on_h}.  If the error $\Delta r$ contains $1\%$ of the observed population, it could easily bias the estimation in the Hubble constant by more than $1\%$. Therefore, to achieve a high-precision cosmology from statistical dark siren, it would require a high level of accuracy in the astrophysical model with fine details captured. 

Note further that our Eq.~(\ref{eq:bias_dr_on_thetac}) applies not only to cosmological parameters but also astrophysical ones as we can simply replace $\vect{\theta}^C$ to $\vect{\theta}^A$, or any other subset of $\vect{\theta}$. This could be of astrophysical significance. For example, the location of the mass gap due to pair instability supernovae could be biased by substructures produced by dynamical formation channels or the redshift-dependence in the mass function~\citep{Mukherjee:21b, MariaEzquiaga:22, Karathanasis:22}. Our Eq.~(\ref{eq:bias_dr_on_thetac}) thus provides a simple and analytical way to quantify the bias. 

As a first step, our current model does not include the statistical error on each individual event's component mass and luminosity distance. This may be a subdominant effect for events that are well above the detection threshold, which are typically the ones selected for population studies (see, e.g., \citealt{LVK:21pop, LVK:21cos, Roulet:21}). 
Intuitively, the uncertainty on each event's parameters slightly blurs the measured distribution and smears out sharp features. Yet since both $p(m_1, m_2)$ and $p(z)$ are smooth functions in our study (and in \citealt{LVK:21cos}), such a blurring should not be significant(, but see the discussion below on galaxy catalogs). However, information of the population is also contained in sources that are marginally detectable (or undetectable; see the discussion in \citealt{Roulet:20}). These events could happen at locations where $P_{\rm det}$ has large derivatives with respect to $\vect{\theta}$ and thus may potentially contribute to the Fisher information. To utilize them properly, incorporating their parameter estimation errors would be critical, and we plan to investigate this in a follow-up study. 
%Maybe fine if sample size is sufficiently large. But to make prediction for the near future, we should incorporate individual event's uncertainty. 

We also assumed the galaxy catalog provides only the smoothed shape of the redshift model $p(z)$. This is the case because the GW event localization accuracy is currently limited. In the other limit where a BBH could be localized to a single host galaxy (which can be achieved with a decihertz space-borne detector; \citealt{Kuns:20}), a dark siren would behave effectively like 
a bright BNS event with EM counterpart identified, because the host galaxy in this case can be identified from the sky localization~\citep{Chen:16, Borhanian:20, Seymour:22}.  This could lead to a strong constraint in cosmology~\citep{Chen:18} without needing assumptions in the underlying population model. 
In the intermediate case, an accurate localization plus a complete galaxy catalog could mean sharp spikes in $p(z)$ and therefore $r(z)$. Whereas $h$ can be nearly degenerate with an overall power-law slope $\gamma$ in $\psi(z)$ (which is also the limiting factor on how well we can measure $h$; Fig.~\ref{fig:GWTC3_fisher_err}), it could hardly be confused with sharp spikes. 
Therefore the constraints on $h$ could thus be improved. 
Besides using the location of each individual event, the spatial clustering of BBH events is yet another possibility to enhance our constrain on cosmology and reduce its systematic errors~\citep{Mukherjee:21, Scelfo:20, CigarranDiaz:22,Mukherjee:22}. 
A more quantitative study incorporating these effects coherently is to be carried out in future investigations. 

\begin{acknowledgements}
We thank Katerina Chatzioannou, Hsin-Yu Chen, and the LVC colleagues for helpful discussions and comments during the preparation of the manuscript.
This material is based upon work supported by NSF's LIGO Laboratory which is a major facility fully funded by the National Science Foundation.
H.Y. acknowledges the support of the Sherman Fairchild Foundation.  Y.C., B.S., and Y.W.\ acknowledge support from the Brinson Foundation, the  Simons Foundation (Award Number 568762), and by NSF Grants PHY-2011961, PHY-2011968, PHY--1836809.

\end{acknowledgements}

\software{\texttt{Python3}~\citep{VanRossum:09}, \texttt{NumPy}~\citep{Harris:20}, \texttt{SciPy}~\citep{Virtanen:20}, \texttt{Matplotlib}~\citep{Hunter:07}, \texttt{PYCBC}~\citep{Nitz:22}.
}

\appendix
\section{Validation of the methodology}
\label{sec:validation}

In this Appendix, we further validate our approach by reproducing some results from \cite{Fishbach:18} and \cite{Farr:19}. 

Following \citet{Fishbach:18}, here we consider a \texttt{Truncated Power Law} mass model given by
\begin{equation}
    p(m_1, m_2|\alpha, M_{\rm max})\propto\frac{m_1^{-\alpha}}{m_1-5\,M_\odot}\mathcal{H}(M_{\rm max}-m_1),
    \label{eq:p_m1m2_FHF18}
\end{equation}
where $\mathcal{H}$ is the Heaviside function, and the existence of an upper mass gap $M_{\rm max}$ is motivated by the pair-instability supernovae~\citep{Fowler:64}. Since our focus here is to reproduce the results of \citet{Fishbach:18}, we use this mass model despite the fact that it is currently unfavored by the latest data~\citep{LVC:21gwtc2pop, LVK:21cos, Roulet:21}. 
The $\psi(z)$ part in the redshift model (Eq.~\ref{eq:p_z}) given by Eq.~(\ref{eq:psi_low_z}).
% \begin{equation}
%     \psi(z) \propto (1+z)^\lambda.
% \end{equation}
We particularly adopt $(\alpha, M_{\rm max}, \lambda)=(1, 40\,M_\odot, 3)$ in our calculation. The $P_{\rm det}$ is computed following Sec.~\ref{sec:bias} (see Eqs.~(\ref{eq:P_det}) and (\ref{eq:rho_char_FHF18})). 

\begin{figure*}
  \centering
  \includegraphics[width=1.9\columnwidth]{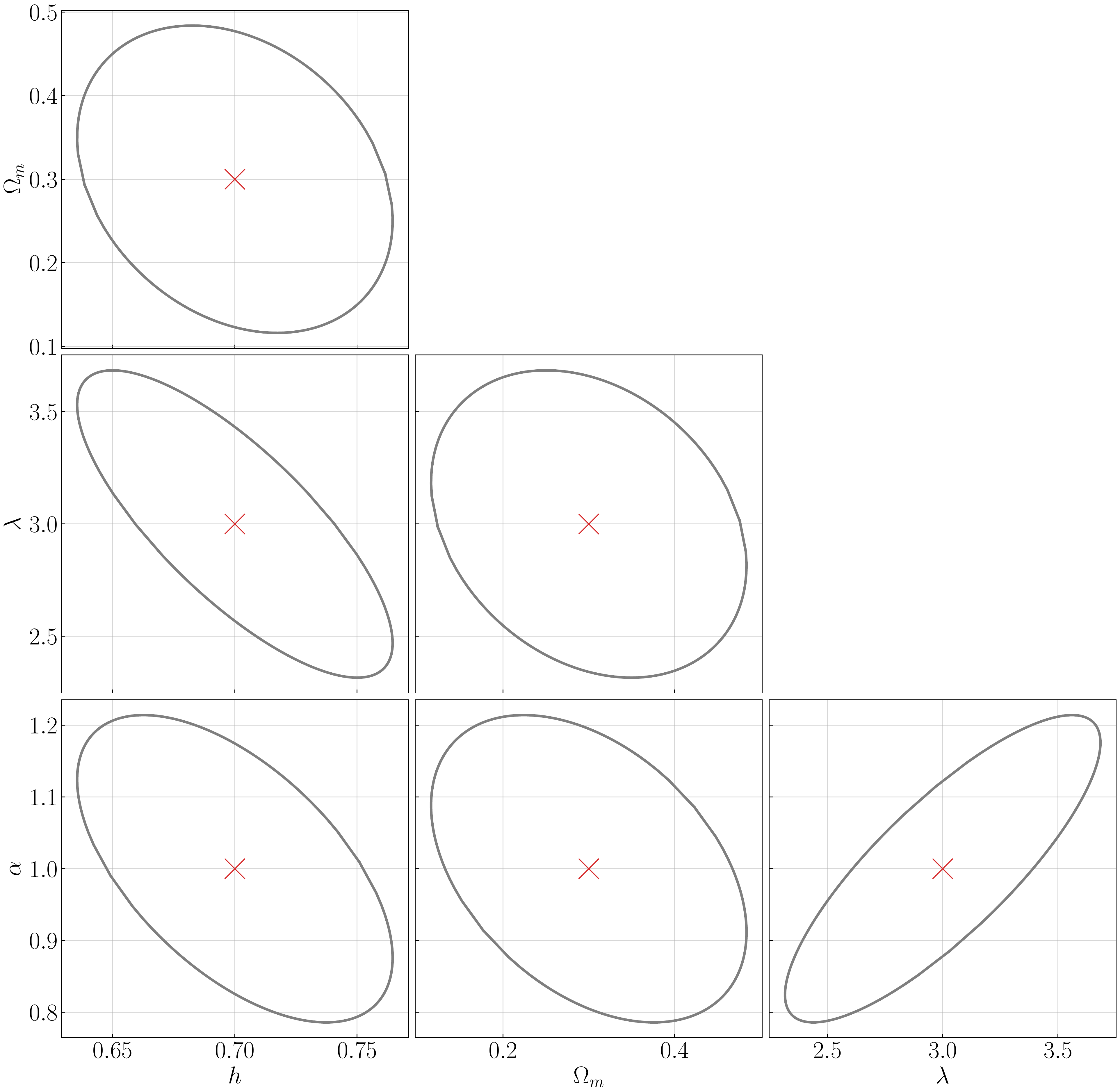}
  \caption{Error ellipses from the 3D $r(m_1, m_2, z)$ distribution assuming 500 BBH events using the model in~\cite{Fishbach:18}. 
  Our results show good agreement with those obtained in \citet{Fishbach:18} and \citet{Farr:19}. 
  It thus validates our approach when $n_{\rm obs}$ is large. 
  %Note that $\Delta \alpha$ and $\Delta \lambda$ reproduces the top left panel in fig. 5 of Ref.~\cite{Fishbach:18}. Because of the sharp cut-off at $m_{\rm max}$ in the source-frame mass distribution, it allows the constrain of $h$ to be as good as $\sim 3\%$, reproducing the result of Ref.~\cite{Farr:19}.  
  }
\label{fig:fisher_err_vs_m1m2z_FHF18_A_al_1_lam_3}
\end{figure*}

In Fig.~\ref{fig:fisher_err_vs_m1m2z_FHF18_A_al_1_lam_3}, we present the $68\%$ credible interval for the key parameters based on the Fisher information matrix, Eq.~(\ref{eq:fisher}), with $n_{\rm obs}=500$. In particular, we highlight the bottom-right corner of Fig.~\ref{fig:fisher_err_vs_m1m2z_FHF18_A_al_1_lam_3} where we show the error ellipse for $(\lambda, \alpha)$. We notice a positive correlation between the two quantities and their uncertainties are, respectively, $\Delta \lambda = 0.68$ and $\Delta \alpha = 0.21$. Both results show nice agreement with the top-left panel in fig. 5 in \citet{Fishbach:18}. Moreover, because in the mass model, Eq.~(\ref{eq:p_m1m2_FHF18}) there is a clear feature set by $M_{\rm max}$, it thus allows the determination of $h$ as proposed in, e.g., \citet{Farr:19} and demonstrated in the left-most column of Fig.~\ref{fig:fisher_err_vs_m1m2z_FHF18_A_al_1_lam_3}. Consistent with \citet{Farr:19}, we note the uncertainty on $h$ from 500 events is $\Delta h=0.065$. 
The consistency between our Fig.~\ref{fig:fisher_err_vs_m1m2z_FHF18_A_al_1_lam_3} and previous studies thus validates our approach in constraining both the astrophysical and cosmological parameters.

% \clearpage
\bibliography{ref}{}
\bibliographystyle{aasjournal}



\end{document}